\documentclass[12pt,preprint,showpacs,tightenlines,preprintnumbers,nofootinbib]{revtex4}
\usepackage{epsf,graphicx}
\usepackage{amsmath}
\usepackage{mathrsfs}

\def\beq{\begin{equation}}
\def\eeq{\end{equation}}
\def\bea{\begin{eqnarray}}
\def\eea{\end{eqnarray}}
\def\beqa{\begin{equation}\begin{array}{l}}
\def\eeqa{\end{array}\end{equation}}
\def\eqlab#1{\label{eq:#1}}
\def\figlab#1{\label{fig:#1}}
\def\tablab#1{\label{tab:#1}}

\def\eref#1{(\ref{eq:#1})}
\def\Eqref#1{Eq.~(\ref{eq:#1})}

\def\Figref#1{Fig.~\ref{fig:#1}}

\def\Tabref#1{Table \ref{tab:#1}}

\def\sla#1{#1 \hspace{-2mm} \slash}
\def\slap{p \hspace{-2mm} \slash}
\def\slad{\partial \hspace{-2.2mm} \slash}

\def\half{\mbox{\small{$\frac{1}{2}$}}}

\def\thalf{\mbox{\small{$\frac{3}{2}$}}}
\def\quarter{\mbox{\small{$\frac{1}{4}$}}}
\def\third{\mbox{\small{$\frac{1}{3}$}}}

\def\barr{\left(\begin{array}{c}}
\def\earr{\end{array}\right)}
\def\bmat{\left(\begin{array}{cc}}
\def\emat{\end{array}\right)}
\def\al{\alpha}
\def\be{\beta}
\def\ga{\gamma} \def\Ga{{\it\Gamma}}
\def\de{\delta} \def\De{\Delta}
\def\veps{\varepsilon}  \def\eps{\epsilon}

\def\la{\lambda} \def\La{{\Lambda}}

\def\si{\sigma} \def\Si{{\it\Sigma}}

\def\pa{\partial}
\def\vrho{\varrho}

\def\pa{\partial}

\def\nn{\nonumber}

\def\rg{{\rm g}}

\def\lag{{\mathcal L}}

\def\mathscr{\mathcal}

\def\3d{3-D}

\def\ol#1{\overline{#1}}

\begin{document}

\title{The nucleon and $\Delta$-resonance 
masses in relativistic chiral effective-field theory }

\author{Vladimir Pascalutsa}
\email{vlad@jlab.org}
\author{Marc Vanderhaeghen}
\email{marcvdh@jlab.org}

\affiliation{Physics Department, The College of William \& Mary, Williamsburg, VA
23187, USA\\
Theory Group, Jefferson Lab, 12000 Jefferson Ave, Newport News, 
VA 23606, USA}

\date{\today}

\begin{abstract}
We study the chiral behavior of the nucleon and $\De$-isobar masses
within a manifestly covariant chiral effective-field theory, consistent
with the analyticity principle.
We compute the $\pi N$ and $\pi\Delta$ one-loop contributions
to the mass and field-normalization constant, and find that
they can be described in terms of universal relativistic loop
functions, multiplied by appropriate spin, isospin and coupling
constants. 
We show that these relativistic one-loop corrections, when properly
renormalized,  obey the chiral power-counting and vanish in the
chiral limit. The results including only the $\pi N$-loop corrections
compare favorably with the lattice QCD data for
the pion-mass dependence of the nucleon and $\De$ masses, 
while inclusion of the $\pi \De$ loops
tends to spoil this agreement.  
\end{abstract}

\pacs{12.39.Fe, 14.20.Dh, 14.20.Gk}%

\preprint{WM-05-124}

\maketitle
\thispagestyle{empty}

The nucleon mass ($M_N\simeq$ 940 MeV)
is much larger than the sum of the masses of its
constituents ($3 m_q\simeq 20$ MeV), hence almost all of it is generated
by the strong interaction among the quarks. An exact
description of this phenomenon has not yet been derived from QCD,
however, tremendous progress has been achieved in 
the numerical computation of the nucleon mass in 
lattice QCD~\cite{DeTar:2004tn,Bernard:2001av}. 
One of the main limitations of the 
lattice QCD studies is that the finite lattice size restricts
the value of quark masses from below, and thus 
the quarks in present lattice studies are much heavier than in reality.
It became a common practice to perform lattice calculations for
different values of quark masses and then {\it extrapolate} the results to the
physical point. 

The extrapolation in the quark mass is not straightforward,
because the non-analytic dependencies, such as $\sqrt{m_q}$
and $\ln m_q$, are shown to be important as one approaches the
small physical value of $m_q$. Therefore naive extrapolations 
often fail, while spectacular non-analytic effects
are found in a number of different quantities, 
see {\it e.g.}, 
Refs.~\cite{Leinweber:2001ui,Hemmert:2003cb,Pascalutsa:2005ts}. 
Fortunately, it is known how
to compute these  non-analytic terms in {\it chiral
effective field theory} (ChEFT) --- a low-energy effective field theory 
of QCD. For recent examples of such calculations for 
the nucleon and other baryon masses see, {\it e.g.}, 
Refs.~\cite{Banerjee:1994bk,Thomas:1999mu,Leinweber:1999ig,Ross05,Bernard:2003xf,Hacker:2005fh}. 
In this Letter we present
a new calculation of the quark-mass dependence of the
nucleon and $\De$-isobar masses in the framework of relativistic
ChEFT, with the emphasis on the {\it analyticity} constraint.

In the ChEFT the interaction is mediated by pions, 
which are the Goldstone bosons of the spontaneously broken
chiral symmetry of QCD. 
The explicit-chiral-symmetry
breaking terms, represented by the pion and quark masses in ChEFT and QCD, 
respectively, are related via the Gell-Mann--Oaks--Renner relation:
$f_\pi^2 m_\pi^2 =- \left<\bar q q\right> 2m_q$, where 
$\left<\bar q q\right> \simeq -(230$ MeV$)^3$ represents the value of the 
quark condensate. Lattice calculations confirm this relation for
a very broad range of quark masses~\cite{Luscher:2005mv}.  
Thus, the quark-mass dependence of quantities in QCD can be translated
to the pion-mass dependence of these quantities in ChEFT and vice
versa. 

As the strength of the Goldstone boson
interactions is proportional to their energy, at sufficiently
low energies a convergent perturbative expansion in ChEFT is possible. 
However, most of the lattice results are presently 
obtained for pion masses above 300 MeV where the chiral
expansion is not expected to converge well. Therefore one resorts
to methods where the leading non-analytic ChEFT results are combined with more
phenomenological techniques such that the resulting approach has a wider
range of applicability \cite{Ross05}, albeit lesser predictive
power. 

Recently it has often been argued~\cite{Becher99,Fucsh03,PHV04} that the manifestly relativistic ChEFT
calculations have, in some cases, 
better convergence than their heavy-baryon (semi-relativistic)
counterparts. This implies that the convergence of the ChEFT expansion is
improved by a resummation of nominally higher-order terms which are
 relativistic corrections to the leading non-analytic terms. 
One can thus improve on the convergence 
of the chiral expansion without loss of predictive power, 
or in plain words, without introducing additional free parameters. 

The original formulation of chiral perturbation theory with nucleons
had been relativistic~\cite{GSS89}, but was claimed
to violate the chiral power counting. The so-called {\it heavy-baryon}
chiral perturbation theory, which treats nucleons semi-relativistically,
was developed to cure the power-counting problem~\cite{Jenkins:1990jv},
and a lot of work has been done since in this direction.
More recently, Becher and Leutwyler~\cite{Becher99} proposed
a manifestly Lorentz-invariant
formulation supplemented with so-called {\it infrared regularization} 
(IR) of loops
in which the chiral power-counting is manifest. At about the same time
it was realized~\cite{Geg99} that 
power-counting can be maintained in a relativistic
formalism without the IR or the heavy-baryon expansions.
The original, straightforward formulation~\cite{GSS89} complies with
chiral power-counting if appropriate renormalizations 
of available low-energy constants are done.   

Power-counting issues apart, the original relativistic formulation 
has a particular advantage over the IR scheme in that it 
preserves {\it analyticity} of the loop contributions~\cite{PHV04}.
The IR procedure spoils the analyticity
 by introducing unphysical cuts in the complex energy
plane. In our work we therefore prefer to use a manifestly Lorentz-covariant
formulation of ChEFT, supplemented with appropriate 
renormalizations~\cite{Geg99}, rather 
than infrared regularizations~\cite{Becher99},
to maintain power counting.


We begin with defining the effective chiral Lagrangian.
Writing here only the first-order terms involving the isovector pseudoscalar
pion field $\pi^a$, the spin-1/2 isospin-1/2
nucleon field $N$ and spin-3/2 isospin-3/2
field $\psi^\mu$ of the $\De$-isobar we have 
(in the conventions of Appendix A):
\bea
\eqlab{lagran}
\lag^{(1)} &=&  
\ol N( i \slad -{M}_{N}) N  + \frac{i g_A}{2 f_\pi M_N} \, 
\ol N \,\ga^{\mu\nu}\ga_5\tau^a (\pa_\mu N)\, \pa_\nu\pi^a \nn\\
&+ &  \ol\psi_\mu (\ga^{\mu\nu\al}i \pa_\al - 
{M}_{\De} \ga^{\mu\nu} ) \,\psi_\nu  + \frac{H_A}{2 f_\pi M_\De}\,
\veps^{\mu\nu\al\la}\, \ol \psi_\mu {\cal T}^a  (\pa_\al \psi_\nu)\,
\pa_\la \pi^a
\\
&+& \frac{i h_A}{2 f_\pi M_\De}\left\{
\ol N\, T^a \,\ga^{\mu\nu\la}\, \pa_\mu \psi_\nu ( \pa_\la \pi^a) 
+ \mbox{H.c.}\right\}  ,\nn
\eea
where $M_\De \simeq 1232$ MeV
is the $\De$-isobar mass, $f_\pi \simeq 92.4$ MeV is the pion
decay constant, $g_A\simeq 1.267$ is the axial coupling
of the nucleon, while $h_A$ and $H_A$
represent 
the lowest order $\pi N \De$ and $\pi \De\De$ couplings, respectively.
In the large-$N_C$ limit they are related to $g_A$ as $h_A=(3/\sqrt{2}) g_A$,
$H_A=9/5 g_A$. The isospin factors 
enter through the Pauli matrices $\tau^a $,
the isospin-1/2-to-3/2 transition matrices $T^a$, and the isospin-3/2
matrices ${\cal T}^a$, with normalizations, $\tau^a \tau^a =3$,
$T^{a\dagger} T^a =2$, ${\cal T}^a {\cal T}^a =5/3$, where summation
over $a$ $(=1,2,3)$ is understood.

To study the chiral behavior of the nucleon and $\De$ masses we
introduce a counter-term Lagrangian containing the corresponding
quantities in the chiral limit $(m_\pi\rightarrow 0$): 
\bea
\eqlab{ctlagran}
\lag^{(1)}_{\rm c.t.} &=&  (M_N -M_N^{(0)})
 \, \ol N N + [Z_{2N}^{(0)} -1] \,\ol N (i\slad -M_N^{(0)} ) N \nn\\
&+&  (M_\De -M_\De^{(0)})\,\ol\psi_\mu \ga^{\mu\nu}  \psi_\nu
+ [Z_{2\De}^{(0)} -1]\, \ol\psi_\mu (\ga^{\mu\nu\al}i \pa_\al - 
{M}_{\De}^{(0)} \ga^{\mu\nu} ) \,\psi_\nu\,, \nn\\
\lag^{(2)}_{\rm c.t.} &=& 4 \,c_{1N}\, m_\pi^2 \, \ol N N - 4\,d_{1N}\, m_\pi^2
\,\ol N (i\slad -M_N^{(0)} ) N \\
&+& 4 \, c_{1\De}\, m_\pi^2\,  \ol\psi_\mu \ga^{\mu\nu}  \psi_\nu
-  4\,d_{1\De}\, m_\pi^2 \ol\psi_\mu (\ga^{\mu\nu\al}i \pa_\al - 
{M}_{\De}^{(0)} \ga^{\mu\nu} ) \,\psi_\nu, \nn
\eea
where $M^{(0)}$ and $Z_2^{(0)}$ represent the chiral-limit value
of the masses and the field-renormalization constants, respectively.

Our choice of the chiral Lagrangian is different from the ones previously
used in the literature 
({\it e.g.},~\cite{Bernard:2003xf,Hacker:2005fh,Becher99})
in two important aspects:
\begin{itemize}
\item[(i)]  The $\pi NN $ coupling differs from the usual {\it pseudovector} 
coupling: 
$
 \frac{g_A}{2 f_\pi} \, 
\ol N \,(\slad \pi^a )\ga_5\tau^a N \,,
$
which is standardly used at this order. 
The difference between this and our $\pi NN$ coupling is of higher order
as can easily be shown by using partial integration and
the Dirac equation for the nucleon field.
Nonetheless, our choice simplifies the calculation 
and, most importantly, allows us
to write down the results for the nucleon and the $\De$ in the same form, see
\Eqref{universalcorr} below.
\item[(ii)]
The couplings of the spin-3/2 field are invariant under a gauge transformation:
\beq
\eqlab{gsym}
\psi_\mu \rightarrow \psi_\mu +\pa_\mu \eps,
\eeq  
with $\eps$ a spinor field. This requirement is called for
by the consistency with the free spin-3/2 field theory~\cite{RaS41}, which
is formulated such that the number of spin degrees of freedom  is constrained
to the physical number, see Refs.~\cite{Pas98,Pas01} for details.
\end{itemize}
Both of these points are crucial for the consistency and elegance of this
calculation. 

Point (ii), in particular, allows us to use
simpler forms for the spin-3/2 propagator. 
Indeed, as can be read off \Eqref{lagran},  
the propagator of the massive spin-3/2 
field is the inverse of the free-field operator:
\beq
\eqlab{free}
(S^{-1})_{\al\be} (p) = \ga_{\al\be\mu}\, p^\mu - m \,\ga_{\al\be} \,,
\eeq
where $p=i\pa$, and $m$ denotes the mass.
However, using the gauge symmetry under \eref{gsym} and 
hence the spin-3/2 constraints:
$\pa\cdot \psi = 0 = \ga\cdot \psi $,
one can obtain other, equivalent, forms of the propagator~\cite{Pa98thesis}.
One can, for example, derive the following gauge-fixing term:
\beq
\lag_{\mathrm{g.f.}} = -i \zeta \left( \pa \cdot \bar\psi \,\, \ga\cdot\psi
-\bar\psi \cdot  \ga\,\, \pa\cdot\psi \right)\,,
\eeq 
with the gauge-fixing parameter $\zeta$, a real number.
Upon adding this term, the free-field operator \Eqref{free} becomes:
\bea
(S^{-1})_{\al\be} (p) & =&  (\slap -m)\, \ga_{\al\be}
+ (1+\zeta) (\ga_\al p_\be - \ga_\be p_\al) \nn\\
&=& \ga_{\al\be}\, (\slap -m) 
- (1-\zeta) (\ga_\al p_\be - \ga_\be p_\al)\,,
\eea
and it is not difficult to find its inverse:
\beq
S^{\al\be} (p) = \frac{\slap+m}{m^2 - p^2}\left[ g^{\al\be} -\third 
\ga^\al \ga^\be + \frac{(1-\zeta)(\zeta \slap +m)}{3(\zeta^2 p^2-m^2)}
(\ga^\al p^\be - \ga^\be p^\al) 
+ \frac{2 (1-\zeta^2)\,p^\al p^\be}{3(\zeta^2 p^2-m^2)} \right].
\eeq
Some simple gauges are:
\bea
\zeta = 1\,: && S^{\al\be} (p)=
\frac{\slap+m}{m^2 - p^2}\left( g^{\al\be} -\third 
\ga^\al \ga^\be \right) \,,\\
\zeta = -1\,: && S^{\al\be} (p)=\left( g^{\al\be} -\third 
\ga^\al \ga^\be \right)
\frac{\slap+m}{m^2 - p^2} \,,\\
\zeta = \infty\,: && S^{\al\be} (p)=
\frac{\slap+m}{m^2 - p^2} {\mathscr P}^{(3/2)\,\al\be}(p),
\eea
where
\beq
{\mathscr P}^{(3/2)\,\al\be}(p) =\frac{2}{3} 
\left(g^{\al\be} - \frac{p^\al p^\be}{p^2} \right) +\frac{\slap}{3 p^2} 
\ga^{\al\be\mu}\, p_\mu 
\eeq
is the covariant 
spin-3/2 projection operator. Obviously, $\zeta =0 $ corresponds
with the usual Rarita-Schwinger propagator. It is interesting to observe
that for $\zeta \neq 0 $ the propagator has a smooth massless limit.
We would like to stress that our results are independent of the
gauge-fixing parameter, because all the spin-3/2 couplings used
here are symmetric with respect to the gauge transformation \eref{gsym}.

In this spin-3/2 formalism the $\De$ self-energy takes a simple form:
\beq
\Si_{\al\be}(p) = \Si(\slap) \,{\mathscr P}^{(3/2)}_{\al\be}(p),
\eeq
where $\Si(\slap)$ has the spin-1/2 Lorentz structure. Thus, both
nucleon and $\De$-isobar self-energies can be expressed in the
same Lorentz form, without complications of the lower-spin
sector of the spin-3/2 theory considered 
in~\cite{Korpa:2004sh,Kaloshin:2004jh}. 

\begin{figure}[tb]
\centerline{  \epsfxsize=7cm
  \epsffile{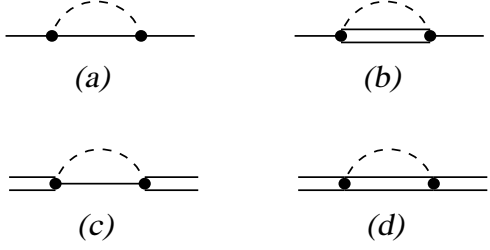} 
}
\caption{The nucleon and $\De$ self-energy contributions
considered in this work. Double lines represent
the $\De$ propagators.}
\figlab{diagrams}
\end{figure}

\begin{table}[tb]
\begin{tabular}{c|c|c}
\hline
\hline
$\,\, {C}_{B}  {}^{B'}$ &  $N$
& $\De$ \\
\hline
$N\,$ & $(3 g_A)^2 $ 
& $ \left( 2 h_A \frac{M_N}{M_\De}\right)^2 $ \\
$\De\, $ & $ h_A^2 $  &   
$ \left(\frac{5}{3} H_A\right) ^2 $ \\
\hline\hline
\end{tabular}
\caption{The coefficient $C_{B B^\prime}$ entering the 
$\pi N$- and $\pi \De$-loop contributions in the baryon mass formula  
\Eqref{universalcorr}. The rational numbers in the brackets represent
the spin (= isospin) factors.}
\tablab{coefs}
\end{table}

Furthermore, in explicit calculations we find
that this form for the nucleon and the $\De$ can be written 
in a universal expression. Namely,  
the one-pion-loop contribution of a baryon $B'$ to the self-energy of a baryon $B$, see \Figref{diagrams}, can generically be written as:
\beq
\Si_B(\slap) = \frac{C_{BB'}}{3(2f_\pi M_B)^2} \frac{1}{i}\!\int \!\frac{d^4 k}{(2\pi)^4} 
\frac{1}{k^2-m_\pi^2} \frac{\slap - \sla{k} + M_{B'}}{(p-k)^2-M_{B'}^2}
\left[p^2 k^2 - (p\cdot k )^2\right]\,,
\eeq
where $C_{BB'}$ is given by the corresponding
coupling constant squared, multiplied by the spin and isospin factors,
see \Tabref{coefs}. To bring the spin-3/2 
$\De$-isobar contributions to this form, the identities
listed in Appendix A are helpful. 
The similarity of
the nucleon and $\De$-isobar loop contributions, pointed out
earlier in Ref.~\cite{Cohen:1992uy}, is thus obtained here
in the formalism of relativistic baryon ChEFT. 

To evaluate the loop integral we use the standard Feynman-parameter trick:
$(AB)^{-1} = \int_0^1 \!dx\, [xA+(1-x)B]^{-2}$, and after the change
of variable $k\rightarrow k+(1-x)p$, obtain: 
\beq
\Si_B(\slap) =\frac{C_{BB'}}{3(2f_\pi M_B)^2 } \int_0^1 \! dx\,(x\slap+M_{B'})
(p^2 g^{\mu\nu} - p^\mu p^\nu)
\frac{1}{i}\!\int \!\frac{d^4 k}{(2\pi)^4} \frac{k_\mu k_\nu}{(k^2-{\cal M}^2)^2},
\eeq
with ${\cal M}^2 = m_\pi^2 x + M_{B'}^2 (1-x) - x(1-x) p^2$.
The latter integral can be computed via dimensional regularization (for
$d \rightarrow 4^- $):
\beq
\frac{1}{i}\!\int \!\frac{d^d k}{(2\pi)^d} \frac{k_\mu k_\nu}{(k^2-{\cal M}^2)^2} = - g_{\mu\nu} \frac{ {\cal M}^2 }{2(4\pi)^2}
\left[ -\frac{2}{4-d} + \ga_E -1 -\ln 4\pi + \ln ({\cal M}^2/\La^2) \right] ,
\eeq
where the Euler constant $\ga_E = - \Gamma'(1) \simeq 0.5772 $,
and $\La$ is a renormalization scale.

Writing the self-energy in general as $\Si(\slap)  = \si(s)+
(\slap-M_B) \tau(s) $,
with $s=p^2$, we find that
\begin{subequations}
\bea
\si(s) &=& -  \frac{C_{BB'} }{2(8\pi f_\pi)^2}\frac{s}{M_B^2}  
\int_0^1 \! dx\,(x M_B+M_{B'}) \, {\cal M}^2 
\left[l_s + \ln ({\cal M}^2/s) \right] ,\\
\tau(s) &=& - \frac{C_{BB'} }{2(8\pi f_\pi)^2 } \frac{s}{M_B^2}
\int_0^1 \! dx\, x  \,{\cal M}^2 
\left[ l_s + \ln ({\cal M}^2/s) \right],
\eea
\end{subequations}
with $l_s=  -2/(4-d) + \ga_E -1 - \ln (4\pi\La^2/s) $. 
Obviously, $\si (M_B^2)$ contributes to the mass of baryon $B$,
while $\tau(M_B^2)$ contributes to its field-renormalization constant (FRC),
namely:
\begin{subequations}
\eqlab{universalcorr}
\bea
M_B &=&
 M_{B}^{(0)} - 4 \, c_{1B}\, m_\pi^2 - \frac{M_B^3}{2(8\pi f_\pi)^2} 
\sum_{B'}\! C_{BB'}\, 
\ol V (\mbox{$\frac{m_\pi}{M_B}$}, \mbox{$\frac{M_{B'}-M_B}{M_B}$}), \\
Z_{2B} &=&
 Z_{2B}^{(0)} - 4 \, d_{1B}\, m_\pi^2 + \frac{M_B^2}{2(8\pi f_\pi)^2}
\sum_{B'}\!  C_{BB'}\, 
\ol W (\mbox{$\frac{m_\pi}{M_B}$}, \mbox{$\frac{M_{B'}-M_B}{M_B}$}),
\eea
\end{subequations}
where  functions $\ol V$ and $\ol W$, given explicitly
in Appendix B, represent
the one-loop contributions with $m_\pi^0$ and $m_\pi^2$ terms subtracted.
The latter terms are subtracted because 
they merely renormalize the available low-energy
parameters, here $M^{(0)}$, $Z_2^{(0)}$, $c_1$, and $d_1$.
It is interesting to note that the loop function $\ol V$
is a relativistic analog (up to a constant factor) of the function $W$ of 
Banerjee and Milana~\cite{Banerjee:1994bk} which represents
the heavy-baryon results.

For $M_B=M_{B'}$ the loop functions simplify considerably:
\begin{subequations}
\bea
\ol V(\mu, 0) & = & \frac{\mu^3}{3} 
\left\{ 8 \left( 1-\quarter \mu^2\right)^{5/2} 
\arccos\frac{\mu}{2} +\frac{\mu}{8}  
\left[ 17 - 2\mu^2 + (30-10\mu^2 +\mu^4) \ln \mu^2  \right]\right\},\\
\ol W(\mu, 0) & = & \frac{\mu^3}{3} 
\left\{ 2 \left( 1-\quarter \mu^2\right)^{3/2} 
\arccos\frac{\mu}{2} +\frac{\mu}{8}  
\left[ 13 - 2\mu^2 + (18-8\mu^2 +\mu^4) \ln \mu^2  \right]\right\} .
\eea
\end{subequations}

\begin{table}[tb]
\begin{tabular}{|c|c|c|c|c|}
\hline
\hline
&  \multicolumn{2}{|c|}{$\pi N$ loop }
& \multicolumn{2}{|c|}{$\pi \De$ loop}  \\
\cline{2-5} 
  & $m_\pi^3$ &  $ \frac{1}{\De} m_\pi^4 \ln m_\pi $
& $m_\pi^3$ &  $ \frac{1}{\De} m_\pi^4 \ln m_\pi  $ \\
\hline
$M_N$ & $-\frac{3}{32\pi f_\pi^2} g_A^2 $ & 0 & $\left\{ \begin{array}{cl}
 0\,,\,\,& \De > m_\pi  \\
 -\frac{1}{24 \pi f_\pi^2} h_A^2, & \De = 0    
\end{array} \right. $ & $\left\{ \begin{array}{cl}
\frac{2}{(8\pi f_\pi)^2} h_A^2 ,& \De > m_\pi  \\
0 \, , & \De = 0    
\end{array} \right. $ \\
$M_\De $ & $ \left\{ \begin{array}{cl}
 0\,,\,\,& \De > m_\pi  \\
 -\frac{1}{96 \pi f_\pi^2} h_A^2, & \De = 0    
\end{array} \right. $ & $ \left\{ \begin{array}{cl}
-\frac{1}{2(8\pi f_\pi)^2} h_A^2 ,& \De > m_\pi  \\
 0 \,, & \De = 0    
\end{array} \right. $ &$-\frac{3}{32\pi f_\pi^2} 
\frac{25}{81} H_A^2 $  & 0   \\
\hline\hline
\end{tabular}
\caption{The leading non-analytic contributions to the nucleon and $\De$ masses
from the $\pi N$- and $\pi \De$-loop.}
\tablab{chiralcoefs}
\end{table}

By studying the expansion of these functions near the chiral limit,
see Appendix B, we find that the chiral expansion for the mass
goes as:
\beq
\eqlab{expansion}
M_B = M_{B}^{(0)} - 4 \, c_{1B}\, m_\pi^2 + \chi_{BB} m_\pi^3 +
\chi_{BB'} \frac{m_\pi^4}{\De} \ln m_\pi + O(m_\pi^4),
\eeq
where $\De = M_\De -M_N > m_\pi $ and
 the chiral coefficients can be read off \Tabref{chiralcoefs}.
This expansion shows explicitly that the (renormalized) relativistic loop
contributions vanish in the chiral limit, as they must. Also, since
there are  no contributions which, near the chiral limit,
scale with positive powers of $\De$, the introduction of counter-terms
of such nature, as is done in \cite{Bernard:2003xf}, is unnecessary,
and would be excessive in this calculation. 

The expansion \eref{expansion} also shows
that the loop contributions obey the chiral power counting. For 
example, in the so-called $\de$-counting~\cite{PP03}, the $\pi N$
and $\pi \De$
loop contributions to the nucleon mass count as $p^3$ and $p^4/\De$,
respectively, which for small $p\sim m_\pi$ agrees with \Eqref{expansion}. 
In the $\de$-counting the one-loop result \Eqref{universalcorr}
thus represents a complete calculation to order $p^4/\De\sim \de^7$.
A full fourth ($p^4$) and $p^5/\De$ order calculation 
of both $\pi N$ and $\pi \Delta$ loops 
requires calculation of diagrams in \Figref{diagrams} 
 with a  $\pi B B^\prime$ vertex from $\lag^{(2)}$
(two derivatives of the pion field) and the tadpole 
contributions. Such a calculation 
is a worthwhile topic for a future work. 

The $\pi N$ contribution to the $\De$ self-energy has an imaginary part
for $m_\pi < \De$, which gives rise to the $\De$ width. 
According to this calculation the width is given by~\cite{PV05}:
\begin{equation}
\Ga_\Delta = \frac{ \pi h_A^2}{12 M_\De^5 (8\pi f_\pi)^2}
[(M_\De+M_N)^2-m_\pi^2]^{5/2} (\De^2-m_\pi^2)^{3/2}.
\end{equation}
The experimental value, $\Ga_\De\simeq 115$ MeV, fixes $h_A\simeq 2.85$,
the value which we shall use in numerical calculations. Note also that
this value is in a much better agreement with the large-$N_C$ value 
$h_A=3 g_A/\sqrt{2}\simeq 2.70$,  than with the SU(6)-relation value, 
$h_A=6 g_A\sqrt{2}/5 \simeq 2.15$.
For the $\De$ axial coupling $H_A$, we use the SU(6) relation,
which in this case coincides with the large-$N_C$ relation: $H_A=(9/5)g_A
\simeq 2.28$.  

\begin{figure}[tb]
\centerline{  \epsfxsize=10cm
  \epsffile{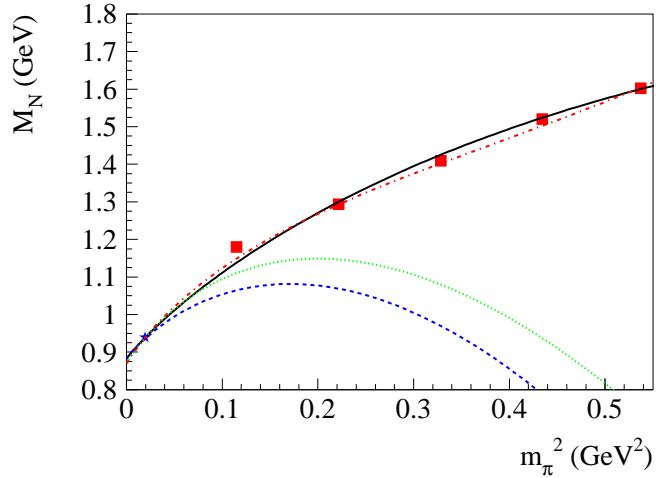} 
}
\caption{(Color online) Pion-mass dependence of the 
nucleon mass. 
The dashed (blue) curve is the leading-nonanalytic
$m_\pi^3$ result, whereas the 
solid (black) curve is the  
full relativistic $\pi N$-loop result, both for parameter values:
$M_N^{(0)} = 0.883$~GeV and $c_{1 N} = -0.87$~GeV$^{-1}$.  
The dotted (green) curve is the
relativistic result for $\pi N + \pi \Delta$ 
loops with $M_N^{(0)} = 0.87$~GeV and 
$c_{1 N} = -1.1$~GeV$^{-1}$. Upon adding to this result
the $m_\pi^4$ term, \Eqref{form4}, with $c_{2 N} = 3.0$~GeV$^{-3}$,
one obtains the dash-dotted (red) curve.  
The  (red) squares are lattice results from the MILC 
Collaboration~\cite{Bernard:2001av}. 
The star represents the physical mass value, which is used in the
fits. }
\figlab{nucmass}
\end{figure}

We are now in position to discuss the numerical results.
\Figref{nucmass} displays the pion-mass 
dependence of the nucleon mass, as given by \Eqref{universalcorr}. 
The  two low-energy constants $M_N^{(0)}$ and $c_{1N}$ are related
to reproduce the physical nucleon mass at the physical pion mass value.
The only free parameter can then be adjusted to reproduce the lattice
data, shown by the squares. Note that these lattice data
are not corrected for finite volume effects, which are known
to increase with decreasing $m_\pi$, and have been estimated
to reach 0.03 GeV for $m_\pi^2=0.1$ GeV$^2$~\cite{Ross05}.
The solid curve in \Figref{nucmass} shows   
the $\pi N$-loop contribution 
to the nucleon mass, with 
$M_N^{(0)} = 0.883$~GeV and $c_{1N} = -0.87$~GeV$^{-1}$.
Thus, the relativistic calculation 
is able to describe the lattice results up to $m_\pi^2 \simeq 0.5$~GeV$^2$ 
with only a single free parameter.
For comparison, the dashed curve shows the corresponding leading non-analytic
result [$m_\pi^3$ term in \Eqref{expansion}]
for the same parameters. One sees that the region of 
applicability of the leading non-analytic term at this order 
is considerably smaller, extends up to 
$m_\pi^2 \simeq 0.05$~GeV$^2$.  
The relativistic result gives a better description out to larger 
pion-mass values due to a  resummation of higher order 
effects ($m_\pi^4 \ln m_\pi$, $m_\pi^5$, etc., terms),
which ensures the correct analyticity properties. 
The pion-nucleon sigma-term can be obtained in this calculation
as $\si_{N} = m_\pi^2 \, d M_N/d m_\pi^2$, taken at physical $m_\pi$:
\beq
\si_N = 67 - 17 = 50 \,\, \left[ \mbox{MeV} \right],
\eeq 
where the first number refers to the contribution of the 
low-energy constant $c_{1N}$, while the second is the 
chiral loop correction.

The dotted curve in \Figref{nucmass} shows the relativistic result 
for both $\pi N$ and $\pi \Delta$ loops according to 
\Eqref{universalcorr}, with slightly re-adjusted low energy 
constants $M_N^{(0)} = 0.870$~GeV, $c_{1N} = -1.1$~GeV$^{-1}$.
One sees that the $\pi \De$ loop gives a substantial contribution 
for larger pion masses and spoils the agreement of our $p^3$ relativistic
calculation with lattice data, above $m_\pi^2 \simeq 0.15 $ GeV$^2$. 
The corresponding sigma-term is
\beq
\si_N = 85 - 17 - 11 = 57 \,\, \left[ \mbox{MeV} \right],
\eeq 
where the numbers refer to the contributions of $c_{1N}$, the $\pi N$, 
and  the  $\pi \De$ loops, respectively.

In absence of a complete fourth order calculation for both $\pi N$ and 
$\pi \Delta$ loop contributions, we estimate the higher order terms here 
in a simple way by allowing for one additional term, proportional to 
$m_\pi^4$ in the baryon mass formula as~:
\beq
\eqlab{form4}
M_B =
 M_{B}^{(0)} - 4 \, c_{1B}\, m_\pi^2 +  c_{2B}\, m_\pi^4 
+ \, \mbox{chiral loops},
\eeq
where the chiral loop contribution is calculated as discussed above 
in \Eqref{universalcorr}. Such a procedure was also proposed before 
in Ref.~\cite{Ross05}, when applying a heavy-baryon formula 
for the non-analytic contribution in the quark mass to lattice results. 
We see from \Figref{nucmass}, that with this 3 parameter formula, 
one can obtain a description of the lattice calculation with the 
relativistic $\pi N + \pi \Delta$ result 
up to $m_\pi^2 \simeq 0.5$~GeV$^2$, using as parameter values~:
$M_N^{(0)} = 0.87$~GeV, $c_{1N} = -1.1$~GeV$^{-1}$, and 
 $c_{2N} = 3.0$~GeV$^{-3}$. 
Although the $m_\pi^4$ term only contributes about 1 MeV to the nucleon mass 
for physical pion mass values, its contribution at $m_\pi^2 = 0.5$~GeV$^2$ 
amounts to about 750 MeV. We notice that in Ref.~\cite{Ross05}  
the value of the coefficient $c_{2N}$ was also found to be large,
which signals the importance 
of further higher-order terms. 

\begin{figure}[tb]
\centerline{  \epsfxsize=10cm
  \epsffile{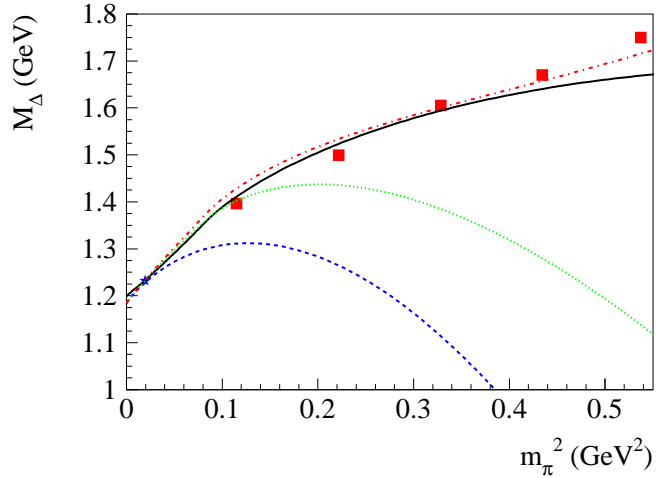} 
}
\caption{(Color online) Pion-mass dependence of the 
$\Delta$ mass. 
The solid (black) curve is the relativistic $\pi N$ loop result, with
$M_\Delta^{(0)} = 1.20$~GeV and $c_{1 \Delta} = -0.40$~GeV$^{-1}$.  
The dashed (blue) curve is the leading non-analytic 
$m_\pi^3$ result (arising from 
$\pi \Delta$ loops), with 
$M_\Delta^{(0)} = 1.185$~GeV and $c_{1 \Delta} = -0.75$~GeV$^{-1}$.  
The dotted (green) curve shows the
relativistic $\pi N + \pi \Delta$ result for the same parameters,
whereas upon adding to this result the $m_\pi^4$ term as in
\Eqref{form4}, with 
$c_{2 \Delta} = 2.0$~GeV$^{-3}$, one obtains the dash-dotted (red) curve.  
The  (red) squares are lattice results from the MILC 
Collaboration~\cite{Bernard:2001av}. 
The star represents the physical mass value, which is used in the fits. }
\figlab{delmass}
\end{figure}

\Figref{delmass} displays the results for the $\Delta$ mass. 
As in the nucleon case, 
the relativistic $\pi N$ loop result, shown by the solid curve,
is able to provide a surprisingly 
good description of the lattice results up to about 
$m_\pi^2 \simeq 0.4$~GeV$^2$, using
$M_\Delta^{(0)} = 1.20$~GeV, $c_{1 \Delta} = -0.40$~GeV$^{-1}$. 
When including the $\pi \Delta$ loops one notices that although the 
convergence of the relativistic calculation (dotted curve in 
\Figref{delmass}) is improved in comparison with the 
leading-nonanalytic result (dashed curve in \Figref{delmass}), 
its agreement with the lattice results is limited to 
$m_\pi^2 \simeq 0.1$~GeV$^2$. 

As for the nucleon, we estimate 
remaining fourth order contributions by the form of \Eqref{form4}. 
Using such a three parameter form, the relativistic $\pi N + \pi \Delta$ 
loop calculation is  able to describe the pion mass dependence 
of the $\Delta$ mass up to $m_\pi^2 \simeq 0.5$~GeV$^2$, 
with 
$M_\Delta^{(0)} = 1.185$~GeV, $c_{1 \Delta} = -0.75$~GeV$^{-1}$, 
and $c_{2 \Delta} = 2$~GeV$^{-3}$. We note that the coefficient 
$c_{2 \Delta}$ is of the same size as for the nucleon and represents a 
500 MeV  mass contribution to the $\Delta$ mass at 
$m_\pi^2 = 0.5$~GeV$^2$. 


In conclusion, we have studied the pion-mass dependence 
of nucleon and $\De$-isobar masses
within the framework of a manifestly covariant chiral effective-field theory. 
We have computed the one-loop $\pi N$ and $\pi \Delta$ graphs 
and obtained a generic expression for those contributions to
the masses and field renormalization constants, \Eqref{universalcorr}. 
We were able to obtain these generic expressions because of
a specific choice of the chiral Lagrangian, where 
the $\pi N \Delta$ and $\pi \Delta \Delta$ 
couplings are constructed to be consistent with
the spin degrees of freedom counting of the relativistic spin-3/2
field. For the $\pi NN$ coupling we adopt a form which is similar
to the above-mentioned $\De$ couplings. 
The resulting relativistic loop corrections obey the chiral power-counting,
after renormalizations of the available counter-terms are done.
The relativistic expressions also contain the nominally higher-order
terms, which are necessary to satisfy the analyticity constraint.
As has been shown here on the example of
the nucleon and $\Delta$ masses,
the convergence of the chiral expansion can be improved in this way,
without introducing additional 
free parameters. 
In particular, we find that the relativistic calculation, including
only the $\pi N$ loops, 
is able to describe the lattice results for both nucleon and $\Delta$ masses 
up to $m_\pi^2 \simeq 0.5$~GeV$^2$ with only one free parameter.
Including the $\pi \Delta$ loops, however, spoils the agreement with 
the lattice result.
We then estimated the effect of the higher-order terms 
by adding  a $m_\pi^4$ term to the baryon mass.
Using the additional free parameter, one is able to 
obtain a description of the lattice calculation with the 
relativistic $\pi N + \pi \Delta$ result 
up to $m_\pi^2 \simeq 0.5$~GeV$^2$.
While a full fourth order calculation of both $\pi N$ and $\pi \Delta$ loops 
is a worthwhile topic for future work, the present relativistic 
chiral-loop calculation 
can be used in the interpolation between full lattice QCD simulations 
and the experimental results.

\appendix
\section{Conventions, rules and identities}

Here we summarize the conventions, Feynman rules, and list a few useful
identities used throughout this work.
\begin{itemize}
\item Conventions: $g^{\mu\nu}=\mbox{diag}(1,-1,-1,-1)$, $\veps^{0123} =1 $,
$\ga_5=i\ga_0\ga_1\ga_2\ga_3$, $(\ga^\mu)^\dagger = \ga^0 \ga^\mu \ga^0$,
$\ga_5^{\dagger} = \ga_5$.
Furthermore,
$\ga$'s denote Dirac's $\ga$-matrices {\it and} their totally-antisymmetric products:
$\ga^{\mu\nu} = \half[\ga^\mu,\ga^\nu]$, 
$\ga^{\mu\nu\al} = \half \{\ga^{\mu\nu},\ga^\al\}$, 
$\ga^{\mu\nu\al\be} = \half [ \ga^{\mu\nu\al},\ga^\be ]$.

\item Propagators:
\begin{subequations}
\bea
S_\pi (p) &=& (p^2 - m_\pi^2 + i\veps)^{-1}, \\
S_N (p) &=& (\slap - M_N + i\veps)^{-1} = (\slap + M_N) \, (p^2 - M_N^2 + i\veps)^{-1}, \\
S_\De^{\al\be} (p) &=& - (\slap + M_\De) \, (p^2 - M_\De^2 + i\veps)^{-1}\,
{\mathscr P}^{(3/2)\al\be}(p).
\eea
\end{subequations}
\item Vertices:
\begin{subequations}
\bea
\Ga_{\pi NN}^{(1)a} (p',p) &=& \frac{g_A}{2 M_N f_\pi} 
 i \veps^{\al \be\vrho\si}
\ga_\al \ga_\be p_\vrho p_\si' \,\tau^a ,\\
\Ga_{\pi N\De}^{(1)a, \,\al} (p',p) 
&=& \frac{h_A}{2 M_\De f_\pi} i \veps^{\al \be\vrho\si}
p_\be' p_\vrho \ga_\si\,\ga_5 T^a, \\
 \Ga_{\pi \De\De}^{(1)a,\,\al\be} (p',p) 
&=& \frac{H_A}{2 M_\De f_\pi} i \veps^{\al \be\vrho\si}
p_\vrho p_\si'  \, {\cal T}^a\,. 
\eea
\end{subequations}
\item Identities:
\begin{subequations}
\bea
&& i \veps^{\mu \nu \vrho \si} \ga_\si \ga_5 =  \ga^{\mu \nu\vrho }
= \half ( \ga^{\mu}\ga^{\nu}\ga^\vrho  - \ga^\vrho\ga^{\nu} \ga^{\mu}),\\
&& i \veps^{\mu \nu \vrho \si} \ga_5 = 
\ga^{\mu \nu \vrho \si} = \quarter ( \ga^\mu \ga^\nu \ga^\vrho \ga^\si
-  \ga^\vrho \ga^\nu \ga^\mu \ga^\si +  \ga^\mu \ga^\vrho \ga^\nu \ga^\si
- \ga^\nu \ga^\si \ga^\mu \ga^\vrho )\,, \\
&&  \veps^\mu_{\,\la\si\al}\,p^{\la}\,\ga^{\si}\, \veps_{\mu\nu\rho\be} 
\,p^{\nu}\,\ga^{\rho}= -p^2 (\rg_{\al\be}+\ga_\al\ga_\be)
+ \slap \ga_\be p_\al  +  \ga_\al p_\be \slap,  \\
&& \veps_{\mu\la\si\al}\,p^{\la}\,\ga^{\si}
\left(\rg^{\mu\mu'}-\third\ga^\mu\ga^{\mu'}\right)
\,\veps_{\mu'\nu\rho\be}  \,p^{\nu}\,\ga^{\rho}= - p^2\,
{\mathscr P}^{3/2}_{\al\be}(p), \\
 &&  \veps^{\al \mu \vrho' \si'} p_{\vrho'} k_{\si'}\, 
\veps^{\be \nu \vrho \si} p_\vrho k_\si \,\,g_{\mu\nu}
= -g^{\al\be} [ p^2 k^2 - (p\cdot k)^2 ] + p^\al p^\be k^2 
+ k^\al k^\be p^2\nn\\ 
&& \hskip5cm - \, p\cdot k \, ( p^\al k^\be + p^\be k^\al) \,, \\
 && \veps^{\al \mu \vrho' \si'} p_{\vrho'} k_{\si'}\, 
\ga_{\mu} \,\ga_{\nu}\,\veps^{\be \nu \vrho \si} p_\vrho k_\si \, {\mathscr P}^{3/2}_{\al\be}(p)
=- p^2\, k^\al k^\be\,{\mathscr P}^{3/2}_{\al\be}(p).
\eea
\end{subequations}
\end{itemize}

\section{Relativistic loop functions} 
Here we explicitly define functions $\ol V$ and $\ol W$ which enter
the mass and FRC correction formula~\eref{universalcorr}. 
\begin{subequations}
\bea
 \ol V(\mu, \de) & \equiv & V(\mu, \de) -  V(0,\de) - V'(0,\de) \mu^2 \, , \\
V(\mu, \de ) & = & \int_0^1
\! dx\, (R+x) \, 
\left\{\mu^2 x + (1-x)(R^2-x) \right\} \, 
\ln\{\mu^2 x + (1-x)(R^2-x) \} \nn \\
&=& \third (R+\al) \left[ \be \, (\mu^2-2\la^2) \,\ln\mu^2
+ \al \, (R^2-2\la^2) \,\ln R^2  -\mbox{$\frac{2}{3}$} (\al^3 + \be^3) \right. \\
&+ & \left.  4\la^2
+ 4\la^4 \,\Omega(\la) \,  \right] 
+\quarter \mu^4 (\ln \mu^2 -\half ) - \quarter R^4 (\ln R^2 -\half )
\,, \nn \\
 V(0,\de) & = & \mbox{$\frac{5}{36}$} + \mbox{$\frac{5}{18}$} R-
\mbox{$\frac{7}{36}$} R^2 - \mbox{$\frac{2}{3}$} R^3 
- \mbox{$\frac{1}{8}$} R^4
+ \mbox{$\frac{1}{6}$} R^5 +\mbox{$ \frac{1}{12}$} R^6  \\
&-&  \mbox{$\frac{1}{6}$} R^5 ( R^3 +2 R^2
-2 R -6) \,\ln R + \mbox{$\frac{1}{12} $}(R^2-1)^3 (1+R)^2 \,\ln|R^2-1|
 \,,\nn \\
V'(0,\de) &\equiv & \left. ({\pa}/{\pa \mu^2}) V(\mu,\de)\right|_{\mu=0} =  
- \mbox{$\frac{1}{18}$} ( 7+9R +3R^2+9R^3+6 R^4) \nn\\
&+&  \third R^5 (3+2R) \, \ln R + \third ( 1+\thalf R -\thalf R^5-R^6)
\,\ln |R^2-1| \,,
\eea
\end{subequations}
where $R\equiv 1+\de$, 
$\be = -\de-\half(\de^2-\mu^2) = \half ( 1-R^2+\mu^2)$, $\al = 1-\be $, 
$\la^2 = \quarter [\de^2-\mu^2][(2+\de)^2-\mu^2]$, and the elementary function
$\Omega$ is defined as:
\beq
\Omega(\la) = \left\{ \begin{array}{lc} 
\frac{1}{2\la} 
\ln \mbox{$\frac{\be-\mu^2-\la}{\be-\mu^2+\la}$}\,, & \la^2 \ge 0\\
-\frac{1}{\sqrt{-\la^2}} 
 \arctan\frac{\sqrt{-\la^2}}{\al\be+\la^2}
\,, & \la^2 < 0 \,.
\end{array} \right.
\eeq

Similarly,
\begin{subequations}
\bea
 \ol W(\mu, \de) & \equiv & W(\mu, \de) -  W(0,\de) - W'(0,\de) \mu^2 \, , \\
W(\mu, \de ) & = & \int_0^1
\! dx\, x \, 
\left\{\mu^2 x + (1-x)(R^2-x) \right\} \, 
\ln\{\mu^2 x + (1-x)(R^2-x) \} \nn\\
&=& \third \al \left[ \be \, (\mu^2-2\la^2) \,\ln\mu^2
+ \al \, (R^2-2\la^2) \,\ln R^2  -\mbox{$\frac{2}{3}$} (\al^3 + \be^3) \right. \\
&+ & \left.  4\la^2
+ 4\la^4 \,\Omega(\la)\,\right] 
+\quarter \mu^4 (\ln \mu^2 -\half ) - \quarter R^4 (\ln R^2 -\half ),
\nn \\
 W(0,\de) & = & \mbox{$\frac{5}{36}$} -
\mbox{$\frac{7}{36}$} R^2 
- \mbox{$\frac{1}{8}$} R^4 +\mbox{$ \frac{1}{12}$} R^6  \nn\\
&-&  \mbox{$\frac{1}{6}$} R^6 ( R^2-2 R) \,\ln R + 
\mbox{$\frac{1}{12} $}(R^2-1)^3 (1+R^2) \,\ln|R^2-1| 
 \,, \\
W'(0,\de) &\equiv & \left. ({\pa}/{\pa \mu^2}) W(\mu,\de)\right|_{\mu=0} =  
- \mbox{$\frac{1}{18}$} ( 7 +3R^2+6 R^4) \nn\\
&+&  \mbox{$\frac{2}{3}$} R^6  \, \ln R + \third ( 1 -R^6)
\,\ln |R^2-1| \,,
\eea
\end{subequations}

It is useful to know the expansion of these functions for small $\mu$:
\begin{subequations}
\bea
\ol V(\mu, \de)&=&\frac{\mu^4}{\de}
\left[-\ln \mu + \quarter R\,( 1+2R^2) - R^5 \ln R + \half (1+R^5) \,
\ln |R^2-1|\right] +O(\mu^5 ),\\
\ol W(\mu, \de)&=&\frac{\mu^4}{\de}
\frac{[-\ln \mu + \quarter R^2(1+2R^2) - R^6 \ln R + 
\half (1+R^6) \,\ln |R^2-1| ]}{1+R}
+O(\mu^5 ).
\eea
\end{subequations}
For $\de =0$, the expansion takes a different form: 
\begin{subequations}
\bea
\ol V(\mu, 0) & = & \frac{4\pi}{3} \mu^3 + \frac{5}{2} \mu^4 
\left(-\quarter + \ln \mu \right)
+ O(\mu^5), \\
\ol W(\mu, 0) & = & \frac{2\pi}{3} \mu^3 + \frac{3}{2}  \mu^4
\left(-\mbox{$\frac{1}{12}$} + \ln \mu \right)
+ O(\mu^5) .
\eea
\end{subequations}

\begin{acknowledgments}
We  thank Ross Young for useful discussions. 
This work is supported in part by DOE grant no.\
DE-FG02-04ER41302 and contract DE-AC05-84ER-40150 under
which SURA
operates Jefferson Lab.  
\end{acknowledgments}

\end{document}